\documentclass[aps,pra,groupedaddress,amsmath,amssymb,tightenlines,notitlepage,nofootinbib,onecolumn,superscriptaddress,11pt]{scrartcl}
\pdfoutput=1

\PassOptionsToPackage{pdftitle={A Complete Characterization of Passive Unitary Normalizable (PUN) Gaussian States}}{hyperref}

\usepackage[title, toc]{appendix}

\usepackage{pifont}
\usepackage{dsfont}
\usepackage{caption}
  \usepackage{url}
 \usepackage{booktabs,multirow}
  \usepackage{amsthm}
  \usepackage{amsmath}
  \usepackage{amsfonts}
  \usepackage{amssymb}
  \usepackage{amscd}
  \usepackage{bm}
  \usepackage[mathscr]{eucal}
  \usepackage{mathtools} 
  \usepackage{enumitem}
  \usepackage{color}
  \usepackage[automark]{scrlayer-scrpage}
  \usepackage{hyperref}
  \hypersetup{%
        colorlinks,
        breaklinks=true,
        plainpages=false,%
        citecolor=teal,
        linkcolor=violet,
        urlcolor=teal,
        bookmarksopen=true,%
        bookmarksnumbered=false,%
        bookmarksdepth=3%
}
  \usepackage{marginnote}
  \usepackage[top=2.5cm, bottom=2.7cm,left=2.5cm, right=2.5cm, marginparwidth=1.8cm]{geometry}
  \usepackage{framed}
  \usepackage[noabbrev]{cleveref} 
  \usepackage[T1]{fontenc} 
 \usepackage[utf8]{inputenc}
 \usepackage{blkarray}

\usepackage{relsize} 
  
  \usepackage{nicematrix}
   \usepackage{tikz}
   \usetikzlibrary{shapes.geometric, arrows.meta, positioning, shadows.blur}
 \usepackage{forest}
 \usepackage{qtree}  
 \usepackage{wrapfig}
   \usepackage{physics}
  \usepackage[most]{tcolorbox}
  \usepackage{soul}
  \usepackage{lipsum}

  \numberwithin{equation}{section}
 
 \allowdisplaybreaks[1] 
  
  \theoremstyle{definition}  
   \newtheorem{defn}{Definition}[section]
   
   \newtheorem{egs}[defn]{Examples}

   \newtheorem{rmk}[defn]{Remark}
   \newtheorem{rmks}[defn]{Remarks}

  \theoremstyle{plain}  
   \newtheorem{thm}[defn]{Theorem}
   \newtheorem{lem}[defn]{Lemma}

  \theoremstyle{remark} 

   \newcommand{\B}[1]{\mathscr{B}({#1})}

   \newcommand{\bs}{\mathbf{s}}
   
   \newcommand{\bu}{\mathbf{u}}
   \newcommand{\bv}{\mathbf{v}}
   
   \newcommand{\m}{\mathbf{m}}
   \newcommand{\w}{\mathbf{w}}
   \newcommand{\x}{\mathbf{x}}
   \newcommand{\y}{\mathbf{y}}
   \newcommand{\z}{\mathbf{z}}

   \newcommand{\GCn}{\Gamma(\mathds{C}^{n})}

   \newcommand{\BC}{\mathds{C}}
   \newcommand{\BR}{\mathds{R}}

   \definecolor{champagne}{rgb}{0.97, 0.91, 0.81}
   \definecolor{almond}{rgb}{0.94, 0.87, 0.8}
   \definecolor{gainsboro}{rgb}{0.86, 0.86, 0.86}
   \newcommand{\Frame}[1]{\begin{tcolorbox}[colframe = white, colback= blue!10!white, enhanced, breakable]#1 \end{tcolorbox}}
   \newcommand{\Frameremark}[1]{\begin{tcolorbox}[colframe =white, colback=blue!5!white, enhanced, breakable]#1 \end{tcolorbox}}

\usepackage{centernot}

 \newcommand{\numberthis}{\refstepcounter{equation}\tag{\theequation}}  
\makeatletter
\newcommand\footnoteref[1]{\protected@xdef\@thefnmark{\ref{#1}}\@footnotemark}
\makeatother

\let\tr\relax 
\DeclareMathOperator{\tr}{Tr}
\DeclareMathOperator{\diag}{Diag}

\fontsize{12}{14}

\begin{document}
\title{A Complete Characterization of Passive Unitary Normalizable (PUN) Gaussian States}
\author{Tiju Cherian John \thanks{Department of Electrical and Computer Engineering, The University of Arizona, Tucson, AZ~85721, USA\\tijucherian@fulbrightmail.org}, Hemant K. Mishra \thanks{School of Electrical and Computer Engineering, Cornell University, Ithaca, New York~14850, USA\\ hemantmishra1124@gmail.com} \thanks{Department of Mathematics and Computing, Indian Institute of Technology (Indian School of Mines) Dhanbad, Jharkhand 826004, India}, and Saikat Guha \thanks{Department of Electrical and Computer Engineering, University of Maryland, College Park, MD~20742, USA\\ saikat@umd.edu}}

\maketitle

\begin{abstract}
We provide a complete characterization of the class of multimode quantum Gaussian states  that can be reduced to a tensor product of thermal states using only a passive unitary operator. We call these states \textit{passive unitary normalizable} (PUN) Gaussian states.  
The characterization of PUN Gaussian states is given in  three different ways: $(i)$ in terms of their covariance matrices, $(ii)$ using gauge-invariance (a special class of Glauber--Sudarshan $p$-functions), and $(iii)$  with respect to the recently obtained $(A,\Lambda)$ parametrization of Gaussian states in [\textit{J.~Math.~Phys.}~62, 022102 (2021)].   In terms of the covariance matrix, our characterization states that an $n$-mode quantum Gaussian state is PUN if and only if its $2n\times 2n$ quantum covariance matrix $S_{\mathds{R}}$ commutes with the standard symplectic matrix $
     \begin{bsmallmatrix}
             0 &I_n  \\
             -I_n & 0
        \end{bsmallmatrix}$,
    with $I_n$ being the 
    $n\times n$ identity matrix.   
    It is well-known that the so-called gauge-invariant Gaussian states are PUN, but whether the converse is true is not known in the literature to the best of our knowledge.
    We establish the converse in affirmation. 
      Lastly, in terms of the $(A,\Lambda)$-parameterization, we show that a Gaussian state with parameters $(A,\Lambda)$ is PUN if and only if $A=0$.
      \end{abstract}

\Frame{
\tableofcontents
}

      \section{Introduction}
      Gaussian states are one of the most important classes of continuous-variable quantum states due to its accessibility in laboratories and an elegant mathematical formalism \cite{Weedbrook2012-zl, Ferraro2005-bh, Adesso2014-sg, Bhat2019-yg, John2019-io}.
      Mathematically, an $n$-mode quantum Gaussian state $\rho$ is completely described by a $2n\times 2n$ real positive definite matrix $S$ and a vector $\m\in \BC^n$, called the covariance matrix and the mean of the state, respectively.
    It is fundamental and well known that for every  $n$-mode quantum Gaussian state $\rho$ there exists a Gaussian unitary operator $\mathcal{U}$  such that 
\begin{align}\label{eq:gaussian-state-decomposition}
    \mathcal{U}\rho\mathcal{U}^\dagger = \bigotimes_{j=1}^n \tau_j ,
\end{align}
where $\tau_j$ is a one-mode thermal state \cite[Section~3.2.4]{Serafini2017-mz}, \cite{Parthasarathy2013-ja}.  
The class of all Gaussian unitary operators, excluding the displacement operators which change only the mean of a state, can be partitioned into two subclasses: $(i)$ passive Gaussian unitary  operators, the operators of linear optics- obtained from \emph{orthosymplectic} matrices acting on the phase space) and $(ii)$  squeezing  operators, the components of nonlinear optics, obtained from non-orthogonal symplectic matrices acting on the phase space. 

Passive linear optics involves components such as beam splitters, phase shifters, and mirrors.   These elements are readily available, relatively robust, and their operation (mixing light fields) is generally considered practically straightforward to implement and align \cite{Bachor2008-xm}.  These operations preserve Gaussianity but cannot generate nonclassicality (like squeezing or entanglement) from classical inputs like thermal or coherent states \cite{Brask2021-su}.  They are essential for manipulating and interfering quantum states, forming the basis of interferometers and linear optical networks.

We say that a Gaussian state is \emph{passive unitary normalizable} (PUN) if the decomposition \eqref{eq:gaussian-state-decomposition} can be achieved by a passive Gaussian unitary operator. The  class of PUN Gaussian states is significantly important from both a theoretical and a practical standpoint \cite{Wang2007-bj, Bachor2008-xm}. The quantum state of the radiation field collected by an astronomical telescope operating at any center wavelength---a multimode phase-insensitive zero-mean thermal state---falls in this PUN class \cite{Dravins2005-bi}. The quantum state of the codeword at the output of an optical-fiber communications channel with loss and thermal noise with a coherent state transmitter is a PUN Gaussian state \cite{Eisert2007-rl, Caruso2008-ac}. PUN states are simple to prepare experimentally, since it involves a passive linear mixing of thermal states of light via a passive scattering medium, mathematically equivalent to a network of linear passive optical elements such as beamsplitters and phase shifters \cite{Kim2002-dv, Rahimi-Keshari2015-sp}. This is in contrast with the squeezed Gaussian states, preparing which requires more difficult-to-realize nonlinear quantum optics \cite{Leonhardt2010-sf}.

The main aim of this paper is to establish four easy-to-check characterizations for verifying whether a given Gaussian state is PUN or not.
Also, we provide further classification of PUN Gaussian states in terms of their Glauber--Sudarshan $p$-functions.
In summary, this paper provides a systematic study of the class of PUN Gaussian states which is not readily available in the literature.

\subsection{Notations and background}

\textbf{Euclidean spaces:}
    For $\alpha \in \mathds{C}$, let $\mathscr{R}(\alpha)$ and $\mathscr{I}(\alpha)$ denote the real and imaginary parts of $\alpha$, respectively. 
    Let $i$ denote the imaginary unit $\sqrt{-1}$.
    The elements of the $n$-dimensional complex Euclidean space $\mathds{C}^n$ are identified by $n$-column vectors with complex entries, which we denote by bold, small alphabets such as $\bu, \bv$, etc.
    For $\bu \in \mathds{C}^n$, let $u_1,\ldots, u_n$ denote the entries of $\bu$.
    We denote by $\bar{\bu}$ and $\bu^\dagger$ the $n$-column and $n$-row vectors, respectively, whose entries are given by the complex-conjugate of the entries of $\bu$; i.e., $\bar{u}_1,\ldots, \bar{u}_n$.
    For $\bu, \bv \in \mathds{C}^n$, $\bu^\dagger \bv \coloneqq \sum_{j=1}^n \bar{u}_jv_j$ is the Euclidean inner product between $\bu$ and $\bv$.
    We denote the Euclidean norm of $\bu$ by $|\bu|\coloneqq \sqrt{\bu^\dagger \bu}$.
    The $n$-dimensional real Euclidean space $\mathds{R}^n$ is a real subspace of $\mathds{C}^n$.
    For $\x \in \mathds{R}^n$, let $\x^T$ denote the $n$-row vector with the same entries as that of $\x$.

\textbf{Matrices:}
    Let $M_n(\mathds{C})$ denote the set of $n \times n$ complex matrices.
    For $\mathbf{x} \in \mathds{C}^n$, we denote by $\diag(\mathbf{x})$ the diagonal matrix with diagonal entries $x_1,\ldots, x_n$. 
    For $A \in M_n(\mathds{C})$, let $A^T, \bar{A}$, and $A^\dagger$ denote the transpose, entrywise complex-conjugate, and adjoint of $A$, respectively.
    We denote by $\mathscr{R} (A)$ and $\mathscr{I}(A)$ the real and imaginary parts of $A$ given by $(A+\bar{A})/{2}$ and $(A-\bar{A})/{(2i)}$, respectively.
    The matrix $A$ is said to be symmetric if $A^T=A$, and $A$ is called unitary if it satisfies $A^\dagger A=I_n$. We denote the set of $n \times n$ unitary matrices by $U_n(\mathds{C})$, also known as the unitary group. 
     The matrix $A$ is called selfadjoint or Hermitian if $A=A^{\dagger}$.
    The matrix $A$ is said to be positive semidefinite if it is Hermitian and satisfies  $ \bu^\dagger A\bu \geq 0$ for all $\bu \in \mathds{C}^n$. 
    An invertible positive semidefinite matrix is called a positive definite matrix. 
    
    Let $M_n(\mathds{R})$ denote the set of $n \times n$ real matrices.
    A matrix $U \in M_n(\mathds{R})$ is said to be orthogonal if $U^TU = I_n$.
    We denote the set of $n \times n$ orthogonal matrices by $O_n(\mathds{R})$, also known as the orthogonal group.
    A matrix $A \in M_n(\mathds{R})$ is said to be positive semidefinite if it is symmetric and satisfies 
    $ \x^TA\x  \geq 0$ for all $\x \in \mathds{R}^n$.
    If $A$ is positive semidefinite and invertible, it is called a positive definite matrix.
    In both, real and complex cases, we write $A \geq 0$ to indicate that $A$ is positive semidefinite, and we write $A>0$ to indicate that $A$ is positive definite.
    
\textbf{The symplectic group:}
        Recall that $J$ is the standard symplectic matrix $\begin{bsmallmatrix}0& I_n\\-I_n & 0 \end{bsmallmatrix}$.
       A matrix $M \in M_{2n}(\mathds{R})$ satisfying $M^TJM=J$ is known as a {symplectic matrix}, and the set of such matrices forms a group under matrix multiplication known as the symplectic group.
       We denote the symplectic group by $Sp_{2n}(\mathds{R})$.
       A subgroup of the symplectic group that plays important role in the exposition of this article is the \emph{orthosymplectic group}, given by the intersection of the orthogonal group $O_{2n}(\mathds{R})$ with the symplectic group $Sp_{2n}(\mathds{R})$.
       We denote the orthosymplectic group by $OSp_{2n}(\mathds{R})$ and call each matrix in the orthosymplectic group an \emph{orthosymplectic matrix}.
       There is a known one-to-one correspondence between the unitary group $U_n(\mathds{C})$ and the orthosymplectic group $OSp_{2n}(\mathds{R})$ given by $U_n(\mathds{C}) \ni U \mapsto U_{\mathds{R}} \in OSp_{2n}(\mathds{R})$.
       See \cite[Section~2.1.2]{De_Gosson2011-ko}. 
       This identification of $U_n(\mathds{C})$ by $OSp_n(\mathds{R})$ shall play a crucial role in this article.

\textbf{Real linear operators on $\mathds{C}^n$:}
    The vectors in the $2n$-dimensional real Hilbert space $\BR^{2n}$ are of the form $\left[\x^T \ \y^T \right]^T$, for $\x, \y \in \mathds{R}^n$.
    This provides a one-to-one correspondence between the sets $\BR^{2n}$ and $\BC^n$ given by $\mathbf{m} \equiv \left[\x^T \ \y^T \right]^T \mapsto \mathbf{m}_{\mathds{R}} \equiv \x + i \y$ for all $\x, \y \in \mathds{R}^n$.
    With respect to this correspondence, $\BC^n$ is a $2n$-dimensional \emph{real} Hilbert space with the corresponding inner product between $\bu, \bv \in \BC^n$ defined by $ \mathscr{R} ( \bu^\dagger\bv)$.
    Let us fix an orthonormal basis of the real Hilbert space $\mathds{C}^n$ given by $\{\mathbf{e}_1,\ldots, \mathbf{e}_n, i\mathbf{e}_1,\ldots, i\mathbf{e}_n\}$, where $\{\mathbf{e}_1,\ldots, \mathbf{e}_n\}$ is the standard basis of $\mathds{R}^n$.
    Let $L: \mathds{C}^n \to \mathds{C}^n$ be a real linear operator; i.e., $L(a \bu + b \bv)= aL\bu+ bL\bv$ for all $a, b \in \mathds{R}$ and $\bu, \bv \in \mathds{C}^n$.
    We denote by $L_\mathds{R} \in M_{2n}(\mathds{R})$ the matrix of the linear operator $L$ with respect to the aforementioned orthonormal basis of $\mathds{C}^n$.
    By definition, there exist $E, F, G, H \in M_n(\mathds{R})$ such that for $\bu = \x+i\y$ with  $\x,\y \in \BR^n$, 
    \begin{equation}\label{eq:L-matrix-representation}
            L\bu = E\x+F\y+i(G\x+H\y),
    \end{equation}
    and
    \begin{equation}\label{eq:L_R}
        L_{\BR} = \bmqty {E&F\\G&H}.
    \end{equation} 
    This gives a one-to-one correspondence between the set of real linear operators $L:\mathds{C}^n \to \mathds{C}^n$ and real matrices $L_\mathds{R}\in M_{2n}(\mathds{R})$.

    The transpose of $L$ is a unique linear operator $L^T: \mathds{C}^n \to \mathds{C}^n$ satisfying for all $\bu, \bv \in \mathds{C}^n$, $\mathscr{R}\left(\bu^\dagger L \bv\right) = \mathscr{R}\left((L^T\bu)^\dagger \bv \right)$.
    The operator $L$ is said to be symmetric if $L=L^T$, and $L$ is called positive semidefinite if it is symmetric, and satisfies $\mathscr{R}\left(\bu^\dagger L \bu\right) \geq 0$ for all $\bu \in \mathds{C}^n$.
    If $L$ is positive semidefinite and invertible, it is called positive definite.
    One can easily verify that $L$ is symmetric, positive semidefinite, and positive definite if and only if its corresponding matrix $L_\mathds{R}$ is symmetric, positive semidefinite, and positive definite, respectively.
    
    A matrix $M \in M_n(\mathds{C})$ can also be viewed as a complex linear operator $M:\mathds{C}^n \to \mathds{C}^n$ given by $\bu \mapsto M\bu$ for all $\bu \in \mathds{C}^n$.
    Hence, the set of $n \times n$ complex matrices forms a ``special class'' of real linear operators in $\mathds{C}^n$ in the sense that they satisfy a stronger condition of complex linearity.
    It is easy to see from \eqref{eq:L-matrix-representation} and \eqref{eq:L_R} that for $M \in M_n(\mathds{C})$, its associated matrix $M_\mathds{R}$ is given by
    \begin{equation}\label{eq:matrix-of-complex-matrix-L} 
        M_{\BR} \coloneqq \bmqty{\mathscr{R} (M)& -\mathscr{I}(M)\\ \mathscr{I}(M)& \phantom{-}\mathscr{R} (M)}.
    \end{equation}
    Matrices in $M_{2n}(\mathds{R})$ of the form \eqref{eq:matrix-of-complex-matrix-L} are precisely those commuting with $J$ as shown in Lemma~\ref{lem:sigma-commutes-j}.
    Henceforth, we shall refer to matrices in $M_n(\mathds{C})$ as linear operators on $\mathds{C}^n$.
    In particular, we refer to unitary matrices as unitary operators.

\textbf{Symplectic operators on $\mathds{C}^n$:} A real linear operator $L:\BC^n\to \BC^n$ that preserves the imaginary part of the Euclidean inner product on $\BC^n$, i.e., for all $\bu, \bv \in \mathds{C}^n$,
    \begin{equation}
        \mathscr{I}( (L\bu)^\dagger L\bv)=\mathscr{I}( \bu^\dagger \bv),
    \end{equation}
    is known as a \textit{symplectic operator}, and we denote the class of all such operators by $Sp_n(\BC)$.
    It is easy to verify that $L$ is a symplectic operator if and only if $L_{\BR} \in Sp_{2n}(\mathds{R})$. 
    See \cite[Section~IV]{John2021-ep}.
    In particular, a unitary operator $U \in U_n(\mathds{C})$ is a symplectic operator and its corresponding matrix $U_{\mathds{R}} \in OSp_{2n}(\mathds{R})$ is orthosymplectic.

\textbf{Williamson's theorem:}
Williamson's theorem \cite{Williamson1936-zu} states that if $S: \mathds{C}^n \to \mathds{C}^n$ is a positive definite real linear operator, then there exists a symplectic operator $L: \mathds{C}^n \to \mathds{C}^n$ such that $L^TSL = D$, where $D$ is a diagonal operator given by $D(\z) =(d_1 z_1,\dots,d_n z_n)$
 for all $\z \in \mathds{C}^n$ and some fixed positive numbers $d_1,\ldots, d_n$. 
 The diagonal operator $D$ is unique up to ordering of $d_1,\ldots, d_n$, and we call it the Williamson's normal form of $S$.
 The numbers $d_1,\ldots, d_n$ are known as the symplectic eigenvalues or Williamson's parameters of $S$.
 In terms of $2n\times 2n$ real matrices, this states that 
 \begin{align}\label{eq:williamson_normal_form_real}
     L_{\BR}^TS_{\BR}L_{\BR} = D_{\mathds{R}}.
 \end{align}
Note that $D_{\mathds{R}}=\diag(d_1,\ldots, d_n) \oplus \diag(d_1,\ldots, d_n)$.
A simple proof of Williamson's theorem in the finite modes can be seen in \cite{Parthasarathy2013-ja} and in the infinite modes in \cite{Bhat2019-yw, John2019-io}.
Recently, an extension of Williamson's theorem to real symmetric matrices was given in \cite{mishra_2024_will_general}.
 Williamson's theorem has attracted much attention of mathematicians and physicists in the past decade and has become a topic of intense study in areas such as matrix analysis \cite{bhatia2015symplectic, HIAI2018129, mishra2020first, bhatia2020schur, bhatia_jain_2021, jain2021sums, jm, paradan2022horn, mishra2023, Son2021-bs, huang2023, son2022symplectic, huang_mishra_2024, mishra2024equality}, symplectic linear algebra \cite{kamat_mishra_2024},  and operator theory \cite{Bhat2019-yw, john2022interlacing, kumar2024approximating}.

    \begin{table}
    \captionsetup{width=0.8\linewidth}
    \caption{\footnotesize Summary of notations and their mathematical definitions.}
    \label{tab:notation}
    \footnotesize
    \centering
    \begin{tabular}
    [c]{|c|l|l|}
    \hline
    Notation & Meaning & Definition \\\hline \hline
    $M_n(\mathds{C})$ & set of $n\times n$ complex matrices &  \\
    $A^T$ & transpose of $A \in M_n(\mathds{C})$ &  \\
    $\bar{A}$ & entry-wise complex conjugate of $A \in M_n(\mathds{C})$ &  \\
    $A^{\dagger}$ & complex conjugate transpose of $A \in M_n(\mathds{C})$ & $\bar{A}^T$ \\
    $\mathscr{R}(\Lambda)$ & real part of $\Lambda \in \mathds{C}^n \cup M_n(\mathds{C})$ & $\frac{1}{2}\left(\Lambda+\bar{\Lambda} \right)$ \\
    $\mathscr{I}(\Lambda)$ & imaginary part of $\Lambda \in \mathds{C}^n \cup M_n(\mathds{C})$ & $\frac{1}{2i}\left(\Lambda-\bar{\Lambda} \right)$ \\
    $I_n$ & the $n \times n$ identity matrix & $\bar{A}^T$ \\
     $U_n(\mathds{C})$ & set of $n \times n$ unitary matrices & $\{U \in M_n(\mathds{C}): U^{\dagger} U = I_n\}$ \\
    $M_n(\mathds{R})$ & set of $n\times n$ real matrices &  \\
     $O_n(\mathds{R})$ & set of $n\times n$ orthogonal matrices &  $\{U \in M_n(\mathds{R}): U^{T} U = I_n\}$ \\
    $J_{2n}$ or $J$ & standard symplectic matrix & $\begin{bsmallmatrix}
        0 & I_n \\ -I_n & 0
    \end{bsmallmatrix}$     \\
    $L_{\mathds{R}}$ & matrix of a real linear operator $L: \mathds{C}^n \to \mathds{C}^n$ &  $L(\x + i\y) \equiv L_{\mathds{R}} \begin{bsmallmatrix} \x \\ \y \end{bsmallmatrix}, \ \forall \x, \y \in \mathds{R}^n$   \\
    $Sp_n(\BC)$& symplectic real linear operators $L:\BC^n \to \BC^n$& $  \mathscr{I}( (L\bu)^\dagger L\bv)=\mathscr{I}( \bu^\dagger \bv), \ \forall \bu, \bv \in \mathds{C}^n$ \\
    $Sp_{2n}(\mathds{R})$&real symplectic group&$\{M \in M_{2n}(\mathds{R}): M^{T}J M = J\}$  \\
    $OSp_{2n}(\mathds{R})$&real orthosymplectic group& $Sp_{2n}(\mathds{R}) \cap O_{2n}(\mathds{R})$  \\
    \hline 
    GS & set of Gaussian states &  \\
    CGS & set of classical Gaussian states &  \\
    PUNGS & set of passive unitary normalizable Gaussian states & \\
    CSGS & set of circularly symmetric Gaussian states &  \\
    \hline
    \end{tabular}
    \end{table}

\subsection{Discussion of the main result}
The three main characterizations of PUN Gaussian states proved in this article are: $(i)$ in terms of their covariance matrices, $(ii)$ using gauge-invariance, and $(iii)$  with respect to the recently obtained $(A,\Lambda)$ parametrization of Gaussian states, where $A$ is an $n \times n$ complex symmetric matrix and $\Lambda$ is an $n \times n$ Hermitian positive semidefinite matrix \cite{John2021-le}.   

In terms of the covariance matrix, we show (Theorem~\ref{thm:S-commute-J}) that an $n$-mode quantum Gaussian state is PUN if and only if its covariance matrix $S_{\BR}$, which is a $2n\times 2n$ real matrix, commutes with the standard symplectic matrix $J\coloneqq \begin{bsmallmatrix}
             0 &I_n  \\
             -I_n & 0
        \end{bsmallmatrix}$ with $I_n$ being the 
    $n\times n$ identity matrix; i.e., 
\begin{align}
    S_{\BR}J = JS_{\BR}.
\end{align}
       This is also equivalent to the fact that $S_{\BR}$ is of the form \[S_{\BR}=\bmqty{\mathscr{R} (X) &-\mathscr{I} (X)\\\mathscr{I} (X)&\phantom{-}\mathscr{R} (X)},\] for some $n\times n$ complex positive definite matrix $X\geq \frac{1}{2}I_n$.
       This follows directly from Lemma~\ref{lem:sigma-commutes-j} combined with the fact that $S_{\mathds{R}}$ satisfies the uncertainty relation $S_{\mathds{R}}\geq \frac{i}{2}J$.

    Coming to the gauge invariance, it is well-known that the so-called mean-zero gauge-invariant Gaussian states are PUN, but whether the converse is true is not known to the best of our knowledge.
    We show that the converse is also true. 
    More precisely, we show that an $n$-mode PUN Gaussian state $\rho$ is a classical state with  its Glauber--Sudarshan $p$-function representation given by \[\rho=\frac{\det(\Sigma)}{\pi^n
      }\int_{\BC^n} \exp{-\bm{\alpha}^{\dagger}\Sigma \bm{\alpha}} \ketbra{\bm{\alpha}}\dd \bm{\alpha}, \]
      for some $n\times n$ complex positive definite  matrix $\Sigma$ (Theorem \ref{thm:PUN-implies-CGS}). This class of gauge-invariant Gaussian states is a proper subclass of classical Gaussian states (CGS) (see Section \ref{sec:classical}). 
      It is easy to see that the covariance matrix $S_{\BR}$ of the CGS we described above is of the form
\[S_{\BR}= \frac{1}{2}I_{2n}+J^T\Sigma_{\BR} J,\]
see Appendix \ref{sec:integral}, and also \cite[Problem 4.2]{Serafini2017-mz}. The fact that any PUN Gaussian state is classical combined with the result that its covariance matrix $S_{\BR}$ commutes with $J$, implies now that the corresponding classical covariance matrix $\Sigma_{\BR}$  commutes with $J$. 
A classification of Gaussian states in terms of its relation with CGS can be seen in Figure \ref{fig:enter-label} in the Appendix.

Lastly, in terms of the parameterization $(A,\Lambda)$ of the Gaussian states obtained in \cite{John2021-le}, we show that a Gaussian state with parameters $(A,\Lambda)$ is PUN if and only if $   A=0$
      (see Theorem \ref{thm: main-result-1}).
Summarizing everything discussed above, we state our main result as follows.
\Frame{
\begin{thm}
Let $\rho$ be a $n$-mode Gaussian state. Then there exists a passive unitary operator $\mathcal{U}$ such that  $\mathcal{U}\rho \mathcal{U}^{\dagger}$ is a displaced thermal state if and only if one of the following four equivalent conditions are satisfied: 
\begin{enumerate}
    \item The covariance matrix $S_{\BR}$  of $\rho$ commutes with $J$.
    \item The covariance matrix $S_{\BR}$ is of the form \[S_{\BR}=\bmqty{\mathscr{R} (X) &-\mathscr{I} (X)\\\mathscr{I} (X)&\phantom{-}\mathscr{R} (X)},\] for some $n\times n$ complex matrix $X\geq \frac{1}{2}I_n$. 
    \item If $(A,\Lambda)$ is the alternate parametrization of $\rho$, then $A = 0$.
    \item The state is a gauge-invariant state.
\end{enumerate}
\end{thm}
}

Note that a PUN Gaussian state is by definition a mean-zero state. So, the above theorem characterizes PUN Gaussian states if we start with a mean-zero Gaussian state.

\subsection{Classification of quantum Gaussian states}

\begin{figure}
    \centering
    \caption{ \footnotesize
    Classification of quantum  Gaussian states
    }
    \label{fig:enter-label}
    \includegraphics[width=0.8\textwidth]{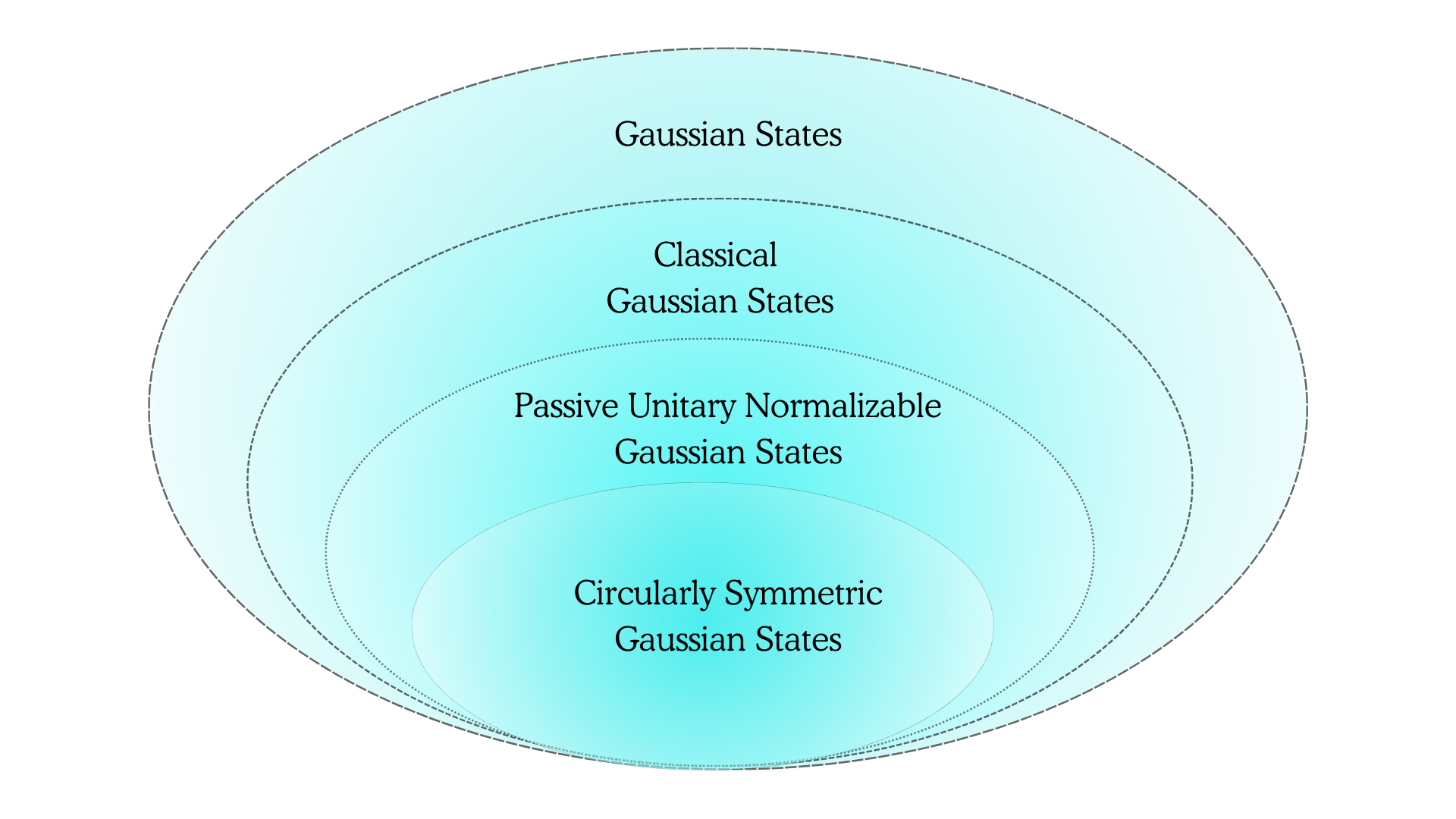}
\end{figure}

 In this section we  describe the different classes of Gaussian states considered in this article.  Figure \ref{fig:enter-label} represents a classification of Gaussian states (GS) on the basis of the structure of their covariance matrices.
     The covariance matrix $S_{\mathds{R}}$ of  a GS satisfies the uncertainty inequality $S_{\mathds{R}}+ \frac{i}{2}J \geq 0$.
    The set of GS contains the set of all classical Gaussian states (CGS).
    The covariance matrix $S_{\mathds{R}}$ of a CGS is of the form $S_{\mathds{R}}=\dfrac{1}{2}I_{2n}+2J^T \Sigma_{\BR} J$, where $\Sigma_{\BR}$ is a real positive semidefinite matrix.
    Inside the set of CGS is the set of PUN Gaussian states.
    The covariance matrix $S_{\mathds{R}}$ of a PUN Gaussian states is of the form $S_{\mathds{R}}=\dfrac{1}{2}I_{2n}+2\tilde{\Sigma}_{\BR}$, where $\tilde{\Sigma}_{\BR}$ is a real positive semidefinite matrix that commutes with $J$.
    The set of PUN Gaussian states has a further subclass of Gaussian states, we call it \emph{circularly symmetric} Gaussian states (CSGS).
    The set of CSGS are precisely those Gaussian states whose covariance matrices are of the form $S_{\mathds{R}}=\frac{1}{2}I_{2n}+2 \begin{bsmallmatrix}
        D & 0 \\
        0 & D
    \end{bsmallmatrix} $, where $D$ is a real diagonal matrix with non-negative diagonal entries.

Table \ref{tab:classical-gs} shows the  $p$ function representation  of different classes of mean-zero classical Gaussian states relevant to our work.

    \begin{table}
    \captionsetup{width=.9\linewidth}
    \footnotesize
    \caption{\footnotesize Classification of classical Gaussian states. Here $\mathbf{N}\equiv \left(N_1^{-1},\dots,N_n^{-1}\right)^T \in \mathds{R}^n$ has non-negative integer entries, $D(\mathbf{N})\coloneqq \diag(\mathbf{N})$, $\bm{\alpha}\equiv \left(\alpha_1,\dots,\alpha_n\right)^T \in \BC^n$, $U \in U_n(\mathds{C})$, and $L\in Sp_n(\mathds{C})$.}
    \label{tab:classical-gs}
     \centering
    \begin{tabular}
    [c]{|c|l|}
    \hline
    Classical GS & $p$-function representation  \\\hline\hline
    & \\
    CSGS &  $\frac{1}{\pi^n
      N_1\cdots N_n}\int_{\BC^n} \exp{-\sum_{j=1}^n\frac{\abs{\alpha_j}^2}{N_j}} \ketbra{\bm{\alpha}}\dd \bm{\alpha}$ 
        \\
    & \\
    PUNGS & $\frac{1}{\pi^n
      N_1\cdots N_n}\int_{\BC^n} \exp{-\left(U\bm{\alpha}\right)^{\dagger}D(\mathbf{N})\left(U\bm{\alpha}\right)} \ketbra{\bm{\alpha}}\dd \bm{\alpha}$ 
       \\ & \\
    CGS &  $\frac{1}{\pi^n
      N_1\cdots N_n}\int_{\BC^n} \exp{-\left(L\bm{\alpha}\right)^{\dagger}D(\mathbf{N})\left(L\bm{\alpha}\right)} \ketbra{\bm{\alpha}}\dd \bm{\alpha}$ \\ & \\
    \hline
    \end{tabular}
    \end{table}

\section{Preliminaries} In this section, we discuss the fundamental objects related to a quantum mechanical system of $n$-modes that we use in this note.  We refer to Sections II and III of \cite{John2021-le} and Chapter 2 of \cite{Parthasarathy1992-tn} for a more detailed discussion on these topics. In addition, we refer to \cite{Androulakis2024-us} for a quick introduction to the thermal states and their displacements. 

Let $\GCn$ denote the boson Fock space over the complex Hilbert space $\BC^n$, and let $\B{\GCn}$ denote the set of bounded linear operators on $\GCn$.  We now summarize the definitions and some properties of quantum-mechanical objects that are relevant to us.
 \begin{enumerate}
 \item  For $\bu\in \BC^n$, the \textit{exponential vector} $\ket{e(\bu)}\in \GCn$ is defined as\begin{equation}
   \ket{e(\bu)}:= 1\oplus\bu\oplus \frac{\bu^{\otimes 2} }{\sqrt{2!}}\oplus \cdots.
 \end{equation} 
 Given $\bu, \bv\in \BC^n$, we have \begin{equation}
     \braket{e(\bu)}{e(\bv)}= \exp{\bu^{\dagger}\bv}.
\end{equation}
\textit{Coherent states} are normalized exponential vectors.  Let $\ket{\bv}$ denote the coherent state at $\bv\in \BC^n$, then \begin{equation}
     \ket{\bv}=  \exp{-\frac{1}{2}\abs{\bv}^2}\ket{e(\bv)}.
 \end{equation}
 The linear span of exponential vectors  forms a dense subspace of $\GCn$.
 \item \textit{The Weyl unitary operator (\textnormal{aka} displacement operator)} at $\bu\in \BC^n$ is denoted by $W(\bu)$, and it is given by 
    \begin{equation}
        W(\bu) \ket{\bv} = \exp{-i \mathscr{I} \left({\bu}^{\dagger}{\bv}\right)}\ket{\bu+\bv}, \quad \forall \bv\in \BC^n. 
     \end{equation}
The Weyl form of the canonical commutation relations is given for $\bu, \bv \in \BC^n$  by
\begin{align}\begin{split}
W(\bu)W(\bv) &= \exp{-i \mathscr{I} \left({\bu}^{\dagger}{\bv}\right)}W(\bu+\bv),\\
W(\bu)W(\bv)& = \exp{-2i\mathscr{I}\left({\bu}^{\dagger}{\bv}\right)}W(\bv)W(\bu). 
\label{eq:2.kb1}
\end{split}
\end{align}
   \item The  field of \textit{fundamental quantum mechanical operators} $\{{p}(\bu)$: $\bu\in \BC^n$\}  are the generators of the strongly continuous one parameter semigroup $t\mapsto W(t\bu), t\in \BR_{+}\cup \{0\}$, of Weyl unitary operators given by
   \[W(t\bu) = e^{-it\sqrt{2}{p}(\bu)}.\] 
   \item Let $\{\mathbf{e}_1, \ldots,\mathbf{e}_n\}$  denote a fixed orthonormal basis of $\BC^n$. The  \textit{momentum, position,  annihilation and creation operators 
} with respect to the given orthonormal basis are obtained as follows:
\begin{align}\label{eq:80}
  {p}_j&:= {p}(\mathbf{e}_j),& {q}_j &:= -{p}(i\mathbf{e}_j),\\
a_j&=\frac{1}{\sqrt{2}}({q}_j+i{p}_j), & a_j^{\dagger}&:= \frac{1}{\sqrt{2}}({q}_j-i{p}_j)
\end{align}
for each $j \in \{1,\ldots, n\}$. The operators ${p}_j$ and ${q}_j$, respectively, are called the $j$-th \emph{\index{momentum}momentum} and \emph{\index{position}position} observables,  $a_j$ and $a_j^{\dagger}$ the $j$-th  \emph{\index{annihilation}annihilation} and \emph{\index{creation}creation} operators  respectively, for all  $j \in \{1,\ldots, n\}$. Then we have the \textit{cononical commutation relations} 
\begin{align}
  \begin{split}\label{ccr-eq:4}
[{q}_j,{q}_k] &=0,~\qquad [{p}_k,{p}_k]=0,\\
[{q}_j,{p}_k]&= i\delta_{jk}I, \quad  1\leq j,k\leq n,
  \end{split}
\end{align}  
where $I$ is the identity operator on $\GCn$.
       \item \label{item:passive-unitary} For any symplectic operator $L$ in the phase space, there exists a unique unitary operator called \textit{the Bogoliubov unitary operator} denoted by $\Gamma(L)$ satisfying for $\bu\in {\BC^n}$,
       \begin{align*} 
        \Gamma(L)W(\bu)\Gamma(L)^\dagger& = W(L\bu),\\
        \mel{\mathbf{0}}{\Gamma(L)}{\mathbf{0}} &>0,
       \end{align*} 
       where $\mathbf{0} \in \mathds{C}^n$ is the zero vector.
       In this case, $\Gamma(L)^{\dagger}=\Gamma(L)^{-1}=\Gamma(L^{-1})$.  For $U \in U_n(\mathds{C})$, the corresponding Bogoliubov unitary operator $\Gamma(U)$ is called the \textit{passive unitary operator} associated with $U$ and satisfies the following additional properties: 
       \begin{align} \Gamma(U)\ket{\bv}&=\ket{U\bv},\label{eq:passive-unitary}\\
       \Gamma(U)^{\dagger} &= \Gamma(U^{\dagger}). 
       \end{align}
       \item A \textit{state} $\rho$ in $\GCn$ is defined as a density operator in $\B{\GCn}$; i.e., $\rho\in \B{\GCn}$, $\rho\geq0$ and $\tr \rho = 1$. The \textit{quantum characteristic function} of a state $\rho$ is the function $\hat{\rho}: \mathds{C}^n \to \mathds{C}$ defined as \begin{equation}
           \hat{\rho}(\bu):= \tr \rho W(\bu), \quad \forall \bu\in \BC^n.
       \end{equation}
    \end{enumerate}

\subsection{Gaussian States} 
\Frameremark{
\begin{defn}\label{defn:Gaussian-state}
An $n$-mode state  $\rho \in \B{\GCn}$  is called a \emph{Gaussian state} (or quantum Gaussian state) if there exists $\m\in \BC^n$ and a real linear, symmetric operator $S$ on $\BC^n$ such that
\begin{equation}\label{eq:Gaussian-qft-1-1}
    \hat{\rho}(\z) = \exp{-2i\mathscr{I}\left({\z}^\dagger{\m}\right)-\mathscr{R}\left( {\z}^\dagger{S}{\z} \right)}, \quad \forall \z \in \BC^n. 
\end{equation}
In this case, we write $\rho = \rho_{\m, S}$, $\m$ is called the \emph{mean annihilation vector}, and $S$ is called the \emph{covariance operator} of the state $\rho$.
\end{defn}
}
If  $\z = \x+i\y$, $\x,\y \in \BR^n$,  (\ref{eq:Gaussian-qft-1-1}) can also be written as 
\begin{align}\label{eq:Gaussian-qft-1}\begin{split}
   \hat{\rho}(\z) & = \exp{-2i\bmqty{\x^T&\y^T}J\bmqty{\mathscr{R} (\m)\\\mathscr{I} (\m)}- \bmqty{\x^T&\y^T}{S_{\BR}}\bmqty{\x\\\y}},    
\end{split}
\end{align}
where recall that $S_\mathds{R}$ is the matrix of $S$ given by \eqref{eq:L_R}. 
It is well known that a $2n\times 2n$ real symmetric matrix $S_{\BR}$ is the covariance matrix of a Gaussian state if and only if $S_{\BR}+\frac{i}{2}J\geq 0$ (see \cite{Parthasarathy2010-su} for an introduction to the mathematical theory of Gaussian states and a  proof of this fact).  We shall use the notations $\rho_{\m_{\BR},S_{\BR}}$ and $\rho_{\m,S_\BR}$  interchangeably to mean the same Gaussian state with the mean vector $\m$ and the covariance operator $S$. 
\Frameremark{
\begin{rmk}\label{rmk:mean-cov-under-passive}
   If $\rho$ is a Gaussian state with mean vector $\m$ and covariance operator $S$ then for every $\bu\in \BC^n$ and symplectic operator $L: \mathds{C}^n \to \mathds{C}^n$, the state $W(\bu)\Gamma(L)\rho \Gamma(L)^{\dagger}W(\bu)^{\dagger}$ is a Gaussian state with mean vector $\bu+L\m$ and covariance operator $(L^{-1})^TSL^{-1}$; i.e., \[W(\bu)\Gamma(L)\rho_{\m,S} \Gamma(L)^{\dagger}W(\bu)^{\dagger} = \rho_{\bu+L(\m), (L^{-1})^TSL^{-1}}.\] In particular, for $U \in U_n(\mathds{C})$, the covariance matrix of the state $W(\bu)\Gamma(U)\rho \Gamma(U)^{\dagger}W(\bu)^{\dagger}$ is $U_{\BR}S_{\BR}U_{\BR}^T$.
\end{rmk}
}
\begin{egs}\label{egs:Gaussian}
    \begin{enumerate}
        \item (\textbf{Vacuum state and coherent states}). The state $\ketbra{\bm{0}}$, $\bm{0}\in\BC^n$ being the zero vector, is an $n$-mode Gaussian state with mean zero and covariance matrix $\frac{1}{2}I_{2n}$, called the $n$-mode vacuum state. For any $\bu\in \BC^n$, the coherent state $\ketbra{\bu}$ is a Gaussian state with mean vector $\bu$ and covariance matrix $\frac{1}{2}I_{2n}$. Moreover, \[\ketbra{\bu} = W(\bu)\ketbra{\bm{0}}W(\bu)^{\dagger}.\]
        \item\label{item:thermal-states} (\textbf{Thermal states and their displacements}). The $1$-mode thermal state with inverse temperature $0<s < \infty$, denoted by $\gamma(s)$, is the state \[\gamma(s)=(1-e^{-s})e^{-sa^{\dagger}a}.\]
        We define $\gamma(\infty) \coloneqq \ketbra{0}{0}$ to be the vacuum state. For an $n$-tuple $\bs = (s_1,\dots,s_n)$ with $0<s_1, \ldots, s_n\leq \infty$, the state \begin{equation}\label{eq:thermal-n-mode}
\gamma(\bs) \coloneqq \otimes_{j=1}^n\gamma(s_j)
        \end{equation} is called the $n$-mode thermal state with inverse temperature $s_j$ in the $j$-th mode. 
     Quantum states of the form $W(\bu)\gamma(\bs)W(\bu)^{\dagger}$, $\bu\in \BC^n$ are called displaced thermal states. 
        Note that 
    displaced thermal states are precisely the Gaussian states with diagonal covariance matrices of the form $D \oplus D$. In this case,   the $n\times n$ diagonal matrix $D$ is given by \[D= \frac{1}{2}\diag\left(\coth \frac{s_1}{2},\dots, \coth \frac{s_n}{2}\right).\]
    See \cite[Proposition 2.12, Corollary 3.6]{Parthasarathy2010-su}, and also  \cite[Proposition 3.4.1]{John2019-io}.
    Now recall from Definition \ref{defn:Gaussian-state} that a covariance operator acts on $\BC^n$ and the corresponding covariance matrix is a $2n\times 2n$ matrix. 
    \end{enumerate}
\end{egs}
\Frameremark{
\begin{rmk}\label{rem:structure_thm_Gaussian_states}
    (\textbf{Structure Theorem for Gaussian States}). Let $\rho_{\m,S}$ be an $n$-mode Gaussian state with mean vector $\m$ and covariance operator $S$. 
    There exists a symplectic operator $L:\BC^n \to \BC^n$ such that $L^TSL=D$ is the Williamson's normal form of $S$ with symplectic eigenvalues $d_1,\ldots, d_n$.
    Then  \[\Gamma(L^{-1})W(-\m)\rho_{\m, S}W(-\m)^{\dagger}\Gamma(L^{-1})^{\dagger} = \gamma(\bs), \] where the vector $\bs$ is given by
    \[d_j=\begin{cases}
        \frac{1}{2}\coth \frac{s_j}{2} & 0< s_j < \infty, \\
        \frac{1}{2} & s_j=\infty.
    \end{cases}\] 
    A simple proof of this structure theorem can be seen in \cite{Parthasarathy2013-ja} for the finite mode case and in \cite{Bhat2019-yg, John2019-io} for the infinite mode case.
\end{rmk}
}
    A Gaussian state can be normalized  to a tensor product of one-mode thermal states by applying a passive unitary followed by a displacement operator if and only if its covariance matrix is orthosymplectically diagonalizable in the sense of Williamson's theorem as stated below.
\Frame{
\begin{lem}\label{lem:orthosymplectic diagnonalization}
 Let $\rho_{\m,S_\BR}$ be a Gaussian state.
 There exists an $n\times n$ unitary matrix $U$ such that $\Gamma(U)\rho\Gamma(U)^{\dagger}$ is a displaced thermal state if and only if the covariance matrix $S_{\BR}$ of the Gaussian state is orthosymplectically  diagonalized by the orthosymplectic matrix $U_{\mathds{R}}$  such that 
    \[U_{\BR}S_{\BR}U_{\BR}^T= \bmqty{D&0\\0&D},\] where $D$ is an $n\times n$ positive diagonal matrix satisfying $D \geq \frac{1}{2}I_n$.
    \end{lem}}
    \begin{proof}
       Immediate from  Remark \ref{rmk:mean-cov-under-passive} and item \ref{item:thermal-states} in Examples \ref{egs:Gaussian} (see also \cite[Theorem 2]{Parthasarathy2013-ja} which show that $D\geq \frac{1}{2}I_n$).
    \end{proof}
\subsection{The \texorpdfstring{$(A, \Lambda)$}{TEXT}-parametrization for Gaussian states}

In this section, we briefly describe the alternative $(A, \Lambda)$-parameterization of quantum Gaussian states introduced in \cite{John2021-le}. Given a symmetric matrix $A$ and a positive semidefinite matrix $\Lambda$ in $M_n(\mathds{C})$, define the $2 n \times 2 n$ matrix
\begin{align}\label{eq:the_M_A_Lambda_matrix}
M(A, \Lambda):=I_{2n}-\left[\begin{array}{cc}
\mathscr{R}(\Lambda) & -\mathscr{I}(\Lambda) \\
\mathscr{I}(\Lambda) &\phantom{-} \mathscr{R}(\Lambda)
\end{array}\right]-2\left[\begin{array}{cc}
\mathscr{R} (A) & \phantom{-}\mathscr{I}(A) \\
\mathscr{I}(A) & -\mathscr{R} (A)
\end{array}\right].
\end{align} 
 
The following result can be concluded from \cite[Theorems V.7, VI.9, VI.12, VI.13, Propositions VI.1, VI.3, Lemmas V.11, VIII.8 ]{John2021-le}. A proof outline of this theorem can be found in Appendix~\ref{app:a-lambda}. It may also be noted that this  $(A,\Lambda)$-parametrization generalizes the $K$-function formalism developed in  \cite{Gagatsos2019-xe}. 
 \Frame{
 \begin{thm}\label{thm:tcj-krp}
    An $n$-mode  state $\rho$ is a Gaussian state if and only if there exists a triple $(\bm{\mu}, A, \Lambda)$, where $\bm{\mu}\in \BC^n$, and $A,\Lambda \in M_n(\mathds{C})$ such that $A$ is symmetric and $\Lambda$ is positive semidefinite satisfying $M(A,\Lambda)>0$. In this case, we have the following.
    \begin{enumerate}
    \item\label{item:1} The generating function $G_\rho(\bu,\bv):=\mel{e(\bar{\bu})}{\rho}{e(\bv)}$ for $\bu, \bv \in \mathds{C}^n$ is given by \begin{equation}\label{eq:Gaussian-generating-function}
        G_\rho(\bu,\bv)=c(A,\Lambda)\exp{ \bu^T\bm{\mu}+ \bar{\bm{\mu}}^T\bv+\bu^TA\bu+ \bu^T\Lambda \bv+ \bv^T\bar{A}\bv},
    \end{equation} where $c(A,\Lambda)\coloneqq \sqrt{\det[M(A,\Lambda)]}.$
    \item\label{item:A-Lambda-under-passive-operation} For $U \in U_n(\mathds{C})$, the parameters of the state $\Gamma(U)\rho\Gamma(U)^{\dagger}$ are given by $\left(U\bm{\mu}, UAU^T,U\Lambda U^{\dagger}\right)$.
        \item\label{item:cov} The covariance matrix $S_{\BR}$ of $\rho$ is related to the pair $(A,\Lambda)$ by 
            \begin{equation}\label{eq:relation_new_old_parameters_covariance}
                \left(\frac{1}{2}I+S_{\BR}\right)^{-1} = M(-A, \Lambda).
            \end{equation}
        \item \label{item:3} The mean vector $\m \in \mathds{C}^n$ of $\rho$ is related to $\bm{\mu}$ by
    \[(I-\Lambda-2AC)\m = \bm{\mu},\]
    where $C:\BC^n\rightarrow\BC^n$ is the complex conjugation map sending $\z\mapsto \bar{\z}$, for all $\z\in \BC^n$. Furthermore, $\m=0$ if and only if $\bm{\mu}=0$.
    \item \label{item:5} The state $\rho$ is  pure if and only if $\Lambda = 0$.
    \item\label{item:7} For every positive semidefinite matrix $\Lambda' \in M_n(\mathds{C})$ satisfying $0\leq \Lambda'\leq \Lambda$, the pair $(A,\Lambda')$ corresponds to a Gaussian state. This provides a continuous path between the pure Gaussian state corresponding to $(A,0)$ and the mixed Gaussian state corresponding to $(A,\Lambda)$. 
     \item \label{item:6} The operator norms of $A$ and $\Lambda$ satisfy \[\norm{A}<\frac{1}{2} \quad\text{and}\quad \norm{\Lambda}< 1.\] 
    \end{enumerate}
\end{thm}
}

\subsection{Classical and gauge-invariant Gaussian states}\label{sec:classical}
\Frameremark{
\begin{defn}
    An $n$-mode quantum Gaussian state $\rho$ is called a \textit{classical Gaussian state} (CGS) if \begin{align*}
\rho&=\int_{\BC^n}\ketbra{\bm{\alpha}}\mathcal{N}_{\bm{\mu},\Sigma}\left(\dd \bm{\alpha}\right)\\
    &=\frac{1}{\sqrt{(2\pi)^{2n}
      \det(\Sigma)}}\int_{\BC^{n}} \exp{-\frac{1}{2}\left(\bm{\alpha}_{\BR}-\bm{\mu}_{\BR}\right)^T\Sigma_{\BR}^{-1}\left(\bm{\alpha}_{\BR}-\bm{\mu}_{\BR}\right)} \ketbra{\bm{\alpha}}\dd \bm{\alpha},
      \end{align*}
where $\mathcal{N}_{\bm{\mu},\Sigma}$ is a $2n$-dimensional  classical normal distribution with mean vector $\bm{\mu}_{\BR}\in \BR^{2n}$ and covariance matrix $\Sigma_{\BR} \in M_{2n}(\BR)$.
\end{defn}
}
The  probability density function $p$ on $\mathds{R}^{2n}$ of $\mathcal{N}_{\bm{\mu},\Sigma}$, is called the Glauber--Sudarshan $p$-function of the CGS given by \[p(\bm{\alpha}_{\BR})=\frac{1}{\sqrt{(2\pi)^{2n}
      \det(\Sigma)}} \exp{-\frac{1}{2}\left(\bm{\alpha}_{\BR}-\bm{\mu}_{\BR}\right)^T\Sigma_{\BR}^{-1}\left(\bm{\alpha}_{\BR}-\bm{\mu}_{\BR}\right)},\quad \bm{\alpha}_{\BR}\in \BR^{2n}\]   \cite{Glauber1963-cw, Glauber1963-ea, Sudarshan1963-be}. It is easy to see that the covariance matrix $S_{\BR}$ of the CGS we described above is of the form
\[S_{\BR}= \frac{1}{2}I_{2n}+J^T\Sigma_{\BR} J.\] See the appendix \ref{sec:integral}, and also \cite[Problem 4.2]{Serafini2017-mz}.
\Frameremark{
\begin{defn}\label{defn:gauge-invariant}
    An $n$-mode Gaussian state is called a \emph{gauge-invariant} state if it is a classical Gaussian state and the covariance matrix  $\Sigma_{\BR}$ of the normal distribution is of the form 
    \begin{align}\label{eq:gauge-inv-cov-matrix}
        \Sigma_{\BR} = \bmqty{\mathscr{R}(K)&-\mathscr{I} (K)\\\mathscr{I} (K)&\mathscr{R}(K)}
    \end{align}
     for some positive definite matrix $K\in M_n(\BC)$.
\end{defn}
}
    If $\rho$ is an $n$-mode gauge-invariant Gaussian state, then it is known to be PUN. In fact, if the covariance matrix of the normal distribution corresponding to the Gaussian state is given by \eqref{eq:gauge-inv-cov-matrix} then a passive unitary operator normalizing the Gaussian state is given by $\Gamma(U)$, where $U$ is an $n \times n$ unitary matrix diagonalizing $K$. 
    However, the converse is not known in the literature to the best of our knowledge.
  We prove in Theorem \ref{thm:PUN-implies-CGS} that PUN Gaussian states are, in fact, gauge-invariant. 
  It should be noted that not all classical Gaussian states are {gauge-invariant}.
  See also \cite[Ch.~V, Sec.~5.II]{Helstrom1969-av} and \cite[Eq.~11]{Holevo1999-pn}.

\section{Main results}
\Frame{
\begin{lem}\label{lem:diagonal-matrices}Let $\rho$ be a Gaussian state with $(A,\Lambda)$ parameters given by Theorem~\ref{thm:tcj-krp}.
Then $\rho$ is a displaced thermal state if and only if both the matrices $A$ and $\Lambda$ are real diagonal matrices.
\end{lem}
}
\begin{proof}
  It is known that an $n$-mode Gaussian state is a displaced thermal state if and only if its covariance matrix is of the form $S_{\mathds{R}} = D \oplus D$ for some $n \times n$ diagonal matrix $D \geq \frac{1}{2}I_n$ (see \ref{item:thermal-states} in Examples \ref{egs:Gaussian}). Using the relation \eqref{eq:relation_new_old_parameters_covariance} of $S_{\mathds{R}}$ with $M(A, \Lambda)$ and the definition \eqref{eq:the_M_A_Lambda_matrix} of $M(A, \Lambda)$, we conclude that $\rho$ is a displaced thermal state if and only if both $A$ and $\Lambda$ are diagonal matrices with real entries.
\end{proof}
\Frame{
\begin{thm}\label{thm: main-result-1}
    Let $\rho$ be an $n$-mode Gaussian state with parameters $(\bm{\mu}, A,\Lambda)$ given in Theorem \ref{thm:tcj-krp}.
    Then there exists an $n\times n$ unitary matrix $U$ such that $\Gamma(U)\rho\Gamma(U)^{\dagger}$ is a displaced thermal state if and only if $A=0$.
\end{thm}
}
\begin{proof} 
Let us assume that $A=0$. Let $U$ be the unitary matrix which diagonalizes $\Lambda$, that is, \[U\Lambda U^{\dagger} = D_{\Lambda},\]
where $D_{\Lambda}$ is a diagonal matrix. It directly follows by item \ref{item:A-Lambda-under-passive-operation} in Theorem~\ref{thm:tcj-krp} and Lemma~\ref{lem:diagonal-matrices} that $\Gamma(U)\rho\Gamma(U)^{\dagger}$ is a displaced thermal state.

    Conversely, let us assume that there exists an $n\times n$ unitary matrix $U$ such that $\Gamma(U)\rho\Gamma(U)^{\dagger}$ is a displaced thermal state.
    Let $S_{\BR}$ be the covariance matrix of $\rho$.
    Now by Lemma~\ref{lem:orthosymplectic diagnonalization}  we have
    \begin{align}\label{eq:orthosym_dia_covariance_matrix}
        U_{\mathds{R}} S_{\BR} U_{\mathds{R}}^T = 
            \begin{bmatrix}
                D_S & 0 \\
                0 & D_S
            \end{bmatrix},
    \end{align}
    where $D_S$ is an $n \times n$ diagonal matrix with positive diagonal entries.
    Furthermore, we also know from item \ref{item:A-Lambda-under-passive-operation} in Theorem~\ref{thm:tcj-krp} and Lemma~\ref{lem:diagonal-matrices} that \begin{align}
        U A U^T &= D_A, \label{eq:diagonalization-A} \\
        U \Lambda U^\dagger &= D_\Lambda,\label{eq:diagonalization-Lambda}
    \end{align}
    where $D_A, D_\Lambda$ are $n \times n$ real diagonal matrices.
    Let $X \coloneqq \mathscr{R}(U)$ and $Y \coloneqq \mathscr{I}(U)$.
    The left-hand side of \eqref{eq:diagonalization-A} can be simplified as
    \begin{align}
        U A U^T     
            &= (X+i Y) (\mathscr{R} (A) + i \mathscr{I}(A))(X^T+i Y^T) \\
            &= (X\mathscr{R}(A) + i X \mathscr{I}(A) + i Y \mathscr{R}(A) - Y \mathscr{I}(A))(X^T+i Y^T)\\     
            &= X\mathscr{R}(A)X^T - Y \mathscr{I}(A)X^T - X \mathscr{I}(A)Y^T - Y \mathscr{R}(A)Y^T \\
            &\hspace{0.5cm} + i \left[  X \mathscr{I}(A)X^T + Y \mathscr{R}(A)X^T + X\mathscr{R}(A)Y^T - Y \mathscr{I}(A)Y^T \right]. \label{eq:real-imaginary-parts-of-uau-transpose}
    \end{align}
    By comparing the real and imaginary parts of \eqref{eq:diagonalization-A} and \eqref{eq:real-imaginary-parts-of-uau-transpose}, we get
    \begin{align}
        X\mathscr{R}(A)X^T - Y \mathscr{I}(A)X^T - X \mathscr{I}(A)Y^T - Y \mathscr{R}(A)Y^T 
            &= D_A, \label{eq:real-part-of-uau-transpose} \\
        X \mathscr{I}(A)X^T + Y \mathscr{R}(A)X^T + X\mathscr{R}(A)Y^T - Y \mathscr{I}(A)Y^T
            &=0. \label{eq:imaginary-part-of-uau-transpose}
    \end{align}
    Similarly, we simplify the left-hand side of \eqref{eq:diagonalization-Lambda} as follows:
    \begin{align}
        U \Lambda U^\dagger    
            &= (X+i Y) (\mathscr{R}(\Lambda) + i \mathscr{I}(\Lambda))(X^T-i Y^T) \\
            &= (X\mathscr{R}(\Lambda) + i X \mathscr{I}(\Lambda) + i Y \mathscr{R}(\Lambda) - Y \mathscr{I}(\Lambda))(X^T-i Y^T)\\ 
            &= X\mathscr{R}(\Lambda)X^T - Y \mathscr{I}(\Lambda)X^T + X \mathscr{I}(\Lambda)Y^T + Y \mathscr{R}(\Lambda)Y^T \\
            &\hspace{0.5cm} + i \left[  X \mathscr{I}(\Lambda)X^T + Y \mathscr{R}(\Lambda)X^T - X\mathscr{R}(\Lambda)Y^T + Y \mathscr{I}(\Lambda)Y^T \right].\label{eq:real-imaginary-parts-of-uLambdau-dagger}
    \end{align}
    Comparing the real and imaginary parts of \eqref{eq:diagonalization-Lambda} and \eqref{eq:real-imaginary-parts-of-uLambdau-dagger}, we get
    \begin{align}
        X\mathscr{R}(\Lambda)X^T - Y \mathscr{I}(\Lambda)X^T + X \mathscr{I}(\Lambda)Y^T + Y \mathscr{R}(\Lambda)Y^T 
            &= D_\Lambda, \label{eq:real-part-of-uLambdau-dagger} \\
        X \mathscr{I}(\Lambda)X^T + Y \mathscr{R}(\Lambda)X^T - X\mathscr{R}(\Lambda)Y^T + Y \mathscr{I}(\Lambda)Y^T 
            &= 0. \label{eq:imaginary-part-of-uLambdau-dagger}
    \end{align}
    
    Combining \eqref{eq:relation_new_old_parameters_covariance} and \eqref{eq:orthosym_dia_covariance_matrix}, we then get
    \begin{align}
        \left(\frac{1}{2}I_{2n}+U_\mathds{R}^T 
            \begin{bmatrix}
                D_S & 0 \\
                0 & D_S
            \end{bmatrix} U_\mathds{R}\right)^{-1} 
            = I_{2n}-
                \left[\begin{array}{cc}
                    \mathscr{R}(\Lambda) & -\mathscr{I}(\Lambda) \\
                    \mathscr{I}(\Lambda) &\phantom{-} \mathscr{R}(\Lambda)
                \end{array}\right]
                +2\left[\begin{array}{cc}
                    \mathscr{R} (A) & \phantom{-}\mathscr{I}(A) \\
                    \mathscr{I}(A) & -\mathscr{R} (A)
                        \end{array}\right],
    \end{align}
    which implies
    \begin{align}\label{eq:covariance_MALambda_orthosymplectic_simplification}
        \left(\frac{1}{2}I_{2n}+
            \begin{bmatrix}
                D_S & 0 \\
                0 & D_S
            \end{bmatrix} \right)^{-1} 
            = I_{2n}-
                U_\mathds{R} \left[\begin{array}{cc}
                    \mathscr{R}(\Lambda) & -\mathscr{I}(\Lambda) \\
                    \mathscr{I}(\Lambda) &\phantom{-} \mathscr{R}(\Lambda)
                \end{array}\right] U_\mathds{R}^T
                +2 U_\mathds{R}\left[\begin{array}{cc}
                    \mathscr{R} (A) & \phantom{-}\mathscr{I}(A) \\
                    \mathscr{I}(A) & -\mathscr{R} (A)
                        \end{array}\right]U_\mathds{R}^T.
    \end{align}
    We simplify the terms on the right-hand side of \eqref{eq:covariance_MALambda_orthosymplectic_simplification} separately as follows. 
    We have
    \begin{align}\label{eq:unitary_orthosyplectic_correspondence}
        U_{\mathds{R}} =
            \begin{bmatrix}
                X & -Y \\
                Y & X
            \end{bmatrix}.
    \end{align}
    Using the representation \eqref{eq:unitary_orthosyplectic_correspondence}, we thus get
    \begin{align}
        &U_\mathds{R} 
        \left[
                \begin{array}{cc}
                    \mathscr{R}(\Lambda) & -\mathscr{I}(\Lambda) \\
                    \mathscr{I}(\Lambda) &\phantom{-} \mathscr{R}(\Lambda)
                \end{array}
        \right] 
        U_\mathds{R}^T \\
                &\hspace{1cm}= 
                    \begin{bmatrix}
                        X & -Y \\
                        Y & X
                    \end{bmatrix}
                    \left[
                    \begin{array}{cc}
                        \mathscr{R}(\Lambda) & -\mathscr{I}(\Lambda) \\
                        \mathscr{I}(\Lambda) & \phantom{-} \mathscr{R}(\Lambda)
                    \end{array}
                    \right]
                    \begin{bmatrix}
                        X^T & Y^T \\
                        -Y^T & X^T
                    \end{bmatrix} \\
                &\hspace{1cm}=
                    \begin{bmatrix}
                        X \mathscr{R}(\Lambda) - Y \mathscr{I}(\Lambda) && -X\mathscr{I}(\Lambda)-Y\mathscr{R}(\Lambda) \\\\
                        X \mathscr{I}(\Lambda) + Y \mathscr{R}(\Lambda) && X \mathscr{R}(\Lambda) - Y \mathscr{I}(\Lambda)
                    \end{bmatrix}
                    \begin{bmatrix}
                        X^T & Y^T \\
                        -Y^T & X^T
                    \end{bmatrix} \\
                &\hspace{1cm}=
                    \begin{bmatrix}
                        \begin{matrix}
                            X \mathscr{R}(\Lambda) X^T - Y \mathscr{I}(\Lambda) X^T \\
                            + X \mathscr{I}(\Lambda) Y^T + Y \mathscr{R}(\Lambda) Y^T
                        \end{matrix} & 
                        \begin{matrix}
                            X \mathscr{R}(\Lambda)Y^T - Y \mathscr{I}(\Lambda)Y^T \\
                            -X\mathscr{I}(\Lambda)X^T-Y\mathscr{R}(\Lambda)X^T
                        \end{matrix} \\\\
                        \begin{matrix}
                            -X \mathscr{R}(\Lambda)Y^T + Y \mathscr{I}(\Lambda)Y^T \\
                            +X\mathscr{I}(\Lambda)X^T+Y\mathscr{R}(\Lambda)X^T
                        \end{matrix} &
                        \begin{matrix}
                            X \mathscr{R}(\Lambda) X^T - Y \mathscr{I}(\Lambda) X^T \\
                            + X \mathscr{I}(\Lambda) Y^T + Y \mathscr{R}(\Lambda) Y^T
                        \end{matrix}
                    \end{bmatrix}.\label{eq:covariance_MALambda_orthosymplectic_simplification-term-one}
    \end{align}
    By substituting the values from \eqref{eq:real-part-of-uLambdau-dagger} and \eqref{eq:imaginary-part-of-uLambdau-dagger} into the matrix \eqref{eq:covariance_MALambda_orthosymplectic_simplification-term-one}, we get
    \begin{align}
        U_\mathds{R} 
        \left[
                \begin{array}{cc}
                    \mathscr{R}(\Lambda) & -\mathscr{I}(\Lambda) \\
                    \mathscr{I}(\Lambda) &\phantom{-} \mathscr{R}(\Lambda)
                \end{array}
        \right] 
        U_\mathds{R}^T
            &= \begin{bmatrix}
                D_\Lambda & 0 \\
                0 & D_\Lambda
            \end{bmatrix}.\label{eq:diagonal-form-ureal-Lambda-ureal-transpose}
    \end{align}
    Similarly, we have
    \begin{align}
        &U_\mathds{R}
        \left[
            \begin{array}{cc}
                    \mathscr{R} (A) & \phantom{-}\mathscr{I}(A) \\
                    \mathscr{I}(A) & -\mathscr{R} (A)
            \end{array}
        \right]
        U_\mathds{R}^T \\
        &\hspace{1cm}=
            \begin{bmatrix}
                X & -Y \\
                Y & \phantom{-} X
            \end{bmatrix}
            \left[
                \begin{array}{cc}
                    \mathscr{R} (A) & \phantom{-} \mathscr{I}(A) \\
                    \mathscr{I}(A) &  -\mathscr{R} (A)
                \end{array}
            \right]
            \begin{bmatrix}
                \phantom{-} X^T & Y^T \\
                -Y^T & X^T
            \end{bmatrix} \\
        &\hspace{1cm} =
            \begin{bmatrix}
                X \mathscr{R} (A)  -Y \mathscr{I}(A) && \phantom{-} X \mathscr{I}(A) + Y \mathscr{R} (A) \\\\
                X \mathscr{I}(A)+Y \mathscr{R} (A)  &&   - X \mathscr{R} (A) + Y \mathscr{I}(A)
            \end{bmatrix}
            \begin{bmatrix}
                \phantom{-} X^T & Y^T \\
                -Y^T & X^T
            \end{bmatrix} \\
        &\hspace{1cm} =
            \begin{bmatrix}
                \begin{matrix}
                   \phantom{-} X \mathscr{R} (A) X^T  -Y \mathscr{I}(A)X^T \\
                     - X \mathscr{I}(A) Y^T - Y \mathscr{R} (A)Y^T
                \end{matrix} &
                \begin{matrix}
                    X \mathscr{R} (A)Y^T  -Y \mathscr{I}(A) Y^T \\
                    X \mathscr{I}(A)X^T + Y \mathscr{R} (A)X^T
                \end{matrix}
                 \\\\
                 \begin{matrix}
                     X \mathscr{R} (A)Y^T  -Y \mathscr{I}(A) Y^T \\
                    X \mathscr{I}(A)X^T + Y \mathscr{R} (A)X^T
                 \end{matrix} &
                 \begin{matrix}
                     - X \mathscr{R} (A) X^T  +Y \mathscr{I}(A)X^T \\
                     + X \mathscr{I}(A) Y^T + Y \mathscr{R} (A)Y^T
                 \end{matrix}
            \end{bmatrix}. \label{eq:covariance_MALambda_orthosymplectic_simplification-term-two}
    \end{align} 
    By substituting the values from \eqref{eq:real-part-of-uau-transpose} and \eqref{eq:imaginary-part-of-uau-transpose} into the matrix \eqref{eq:covariance_MALambda_orthosymplectic_simplification-term-two}, we get
    \begin{align}
        U_\mathds{R} 
        \left[
                \begin{array}{cc}
                    \mathscr{R}(\Lambda) & -\mathscr{I}(\Lambda) \\
                    \mathscr{I}(\Lambda) &\phantom{-} \mathscr{R}(\Lambda)
                \end{array}
        \right] 
        U_\mathds{R}^T
            &= \begin{bmatrix}
                D_A & 0 \\
                0 & -D_A
            \end{bmatrix}.\label{eq:diagonal-form-ureal-a-ureal-transpose}
    \end{align}
    Substitute \eqref{eq:diagonal-form-ureal-Lambda-ureal-transpose} and \eqref{eq:diagonal-form-ureal-a-ureal-transpose} into \eqref{eq:covariance_MALambda_orthosymplectic_simplification} to get
    \begin{align}
            \begin{bmatrix}
                \left(\frac{1}{2}I_n + D_S \right)^{-1} & 0 \\
                0 & \left(\frac{1}{2}I_n + D_S \right)^{-1}
            \end{bmatrix}
            &= 
            \begin{bmatrix}
                I_n-D_\Lambda-2D_A & 0 \\
                0 & I_n-D_\Lambda+2D_A
            \end{bmatrix}.
    \end{align}
    The above equality holds precisely when $D_A=0$.
    From \eqref{eq:diagonalization-A} we thus conclude that $A=0$.
\end{proof}
\Frame{
\begin{thm}\label{thm:S-commute-J}
    Let $\rho$ be an $n$-mode Gaussian state with covariance matrix $S_{\BR}$. Then there exists an $n\times n$ unitary matrix $U$ such that $\Gamma(U)\rho\Gamma(U)^{\dagger}$ is a displaced thermal state if and only if $S_{\BR}$ commutes with $J$, that is \[S_{\BR}J=JS_{\BR}.\]
\end{thm}
}
\begin{proof}
    Observe that $S_{\BR}$ commutes with $J$ if and only if $(\frac{1}{2}I_{2n}+S_{\BR})^{-1}$ commutes with $J$. Let $(\bm{\mu}, A, \Lambda)$ be the parameters of $\rho$ given by Theorem~\ref{thm:tcj-krp}. It directly follows from \eqref{eq:relation_new_old_parameters_covariance} that $S_{\BR}$ commutes with $J$ if and only if $M(-A, \Lambda)$ commutes with $J$.
    It is easy to verify that  $J^TM(-A,\Lambda)J = M(A,\Lambda)$ (see \cite[Lemma VI.2]{John2021-le}). 
    Therefore, $J$ commutes with $S_{\BR}$ if and only if \[M(-A, \Lambda) = M(A, \Lambda). \]
    The above equality holds precisely when $A=0$, which is a direct consequence of the definition \eqref{eq:the_M_A_Lambda_matrix} of $M(A,\Lambda)$.
    The claim thus follows from Theorem \ref{thm: main-result-1} and Lemma~\ref{lem:orthosymplectic diagnonalization}.
\end{proof}
\Frameremark{
\begin{rmk}
    Theorem~\ref{thm:S-commute-J} and Lemma~\ref{lem:orthosymplectic diagnonalization} imply that the covariance matrix of a Gaussian state is diagonalizable by an orthosymplectic matrix in the sense of Williamson's theorem if and only if the covariance matrix commutes with $J$.
    In fact, a more general statement is true: a $2n \times 2n$ real symmetric positive definite matrix $A$ is diagonalizable by an orthosymplectic matrix in the sense of Williamson's theorem if and only if $A$ commutes with $J$, see \cite[Proposition~3.7]{Son2021-bs}.
    The arguments presented in the proof of Theorem~\ref{thm: main-result-1} (from \eqref{eq:orthosym_dia_covariance_matrix} onward) simply work for arbitrary $2n \times 2n$ real symmetric positive definite matrices in place of $S_{\mathds{R}}$.
    Our work thus also provides an alternate proof of the necessary and sufficient condition for orthosymplectic diagonalization in the sense of Williamson's theorem.
\end{rmk}
}
The following lemma is a structure theorem for $2n \times 2n$ real matrices that commute with $J$.
\Frame{
\begin{lem}\label{lem:sigma-commutes-j}
    A matrix $M\in M_{2n}(\BR)$ commutes with $J$ if and only if it is of the form
        $\begin{bsmallmatrix} 
            \mathscr{R} (K)& -\mathscr{I} (K)\\ \mathscr{I}(K)& \phantom{-}\mathscr{R} (K)     
        \end{bsmallmatrix}$
    for some $K \in M_n(\mathds{C})$. 
\end{lem}
}
\begin{proof}
    Let $M \in M_{2n}(\BR)$ and write it in the block form as
    \begin{align}\label{eq:block-form-M}
    M=\begin{bmatrix}
        E & F \\
        G & H
    \end{bmatrix},
    \end{align}
    where $E,F, G, H \in M_{n}(\BR)$.
    It is straightforward to see that $JM=MJ$ if and only if $G=-F$ and $E=H$. 
    Choose $K=E-iF \in M_{n}(\BC)$.
    From the block representation \eqref{eq:block-form-M}, we have $G=-F$ and $E=H$ if and only if $M=\begin{bsmallmatrix} 
            \mathscr{R} (K)& -\mathscr{I} (K)\\ \mathscr{I}(K)& \phantom{-}\mathscr{R} (K)     
        \end{bsmallmatrix}$.
    This completes the proof.

\end{proof}
\Frameremark{
\begin{rmk}\label{rmk:gauge-covriance}
    In the light of the previous lemma and  the definition of gauge-invariant Gaussian states, we see that the classical covariance matrix $\Sigma_{\BR}$ associated with the gauge-invariant state (see Definition \ref{defn:gauge-invariant}) 
 commutes with $J$.
\end{rmk}
}
\Frame{
\begin{thm}\label{thm:PUN-implies-CGS} Let $\rho=\rho_{\m,S}$ be an $n$-mode Gaussian state with covariance operator $S$. Then there exists an $n\times n$ unitary matrix $U$ such that $\Gamma(U)\rho\Gamma(U)^{\dagger}$ is a displaced thermal state if and only if
   $\rho$ is a gauge-invariant Gaussian state.
\end{thm}
}

\begin{proof}
Assume that  $\rho_{\m, S}$ is a gauge-invariant state.
By Lemma~\ref{lem:sigma-commutes-j}, $\Sigma_{\BR}$ commute with $J$, where $\Sigma_{\BR}$ is as in Definition \ref{defn:gauge-invariant}. Since $S_{\BR}=\frac{1}{2}I_{2n}+J^T\Sigma_{\BR} J$ by Theorem \ref{thm:CGS}, we also have that $S_{\BR}$ commutes with $J$. Now, by Theorem \ref{thm:S-commute-J}, we see that there exists  an $n\times n$ unitary matrix $U$ such that $\Gamma(U)\rho\Gamma(U)^{\dagger}$ is a displaced thermal state.

Conversely, assume that there exists an $n\times n$ unitary matrix $U$ such that $\Gamma(U)\rho\Gamma(U)^{\dagger}$ is a displaced thermal state.
First, we show that $\rho_{\m,S}$ is a classical Gaussian state. To this end,  by Theorem \ref{thm:CGS}, it suffices to show that the covariance matrix $S_{\BR}$ satisfies the inequality $S_{\BR}\geq \frac{1}{2}I_{2n}$. 
    Recall from Lemma~\ref{lem:orthosymplectic diagnonalization} that $S_{\BR}$ is orthosymplectically diagonalizable in the sense of Williamson's theorem (see \eqref{eq:orthosym_dia_covariance_matrix}), that is, there exists an orthosymplectic matrix $U_{\BR} \in O_{2n}(\mathds{R})\cap Sp_{2n}(\mathds{R})$ such that
    \begin{align}\label{eq:diagonalization_S_williamson}
        U_\BR ^T S_{\BR} U_{\BR} 
        = 
        \begin{bmatrix}
            D & 0 \\
            0 & D
        \end{bmatrix},
    \end{align}
    where $D$ is an $n \times n$ diagonal matrix such that $D\geq \frac{1}{2}I_n$.
    Consequently, we have  $U_{\BR}^TS_{\BR}U_{\BR} \geq \frac{1}{2}I_{2n}$.
    Since $U$ is an orthogonal matrix, we thus get $S_{\BR} \geq \frac{1}{2}I_{2n}$. Thus, $\rho_{\m, S}$ is a classical state, and there exists a normal distribution $\mathcal{N}_{\m,\Sigma_{\BR}}$ such that  \begin{align*}
        \rho(\m,S)= \int_{\BC^n}\ketbra{\bm{\alpha}}\mathcal{N}_{\m,\Sigma_{\BR}}\left(\dd \bm{\alpha}\right). 
    \end{align*}
    Hence by Theorem \ref{thm:CGS}\[S_\BR=\frac{1}{2}I_{2n} +J^T\Sigma_{\BR} J.\]
    Since $S_{\BR}$ commutes with $J$ by Theorem \ref{thm:S-commute-J} and the fact that $J^T=-J$, we see that $\Sigma$ commutes with $J$. Hence by Lemma~\ref{lem:sigma-commutes-j} we have $\rho_{\m, S}$ is a gauge-invariant state.
\end{proof}
\Frameremark{
\begin{rmk}
    The set of PUN Gaussian states form a proper subset of the set of classical Gaussian states.
    It is simply because any Gaussian state with covariance matrix of the form $S_{\mathds{R}}=\frac{1}{2}I_{2n}+2 J^T \Sigma_{\BR} J$ for $\Sigma_{\BR} \geq 0$ such that $J\Sigma_{\BR} \neq \Sigma_{\BR} J$ is a classical Gaussian state that is not passive unitary normalizable.
    An example of this is $\Sigma_{\BR}= \left[\begin{smallmatrix} 2 & 1 \\ 1 & 1  \end{smallmatrix} \right]$.
\end{rmk}
}
 \begin{appendices}
 \section{Alternate parametrization of Gaussian states}\label{app:a-lambda}

 In this section, we provide an outline of a proof  of Theorem \ref{thm:tcj-krp}.
\begin{proof}
    The main statement of the theorem is exactly the same as that of \cite[Theorem VI.13]{John2021-le}. Note that the proofs of \cite[Propositions V.4, V.5]{John2021-le}, together prove \ref{item:1}. The statements \ref{item:cov}, \ref{item:3}, \ref{item:5} and \ref{item:7} are, respectively, the same as \cite[Proposition VI.3.1, VI.3.2, VI.9.1 and Corollary VI.14]{John2021-le}. To prove \ref{item:6}, first note from \cite[Theorem VI.12]{John2021-le} that the matrix $A$ associated with a pure Gaussian state satisfies $\norm{A}< \frac{1}{2}$. Now by \ref{item:7} we see that a pair $(A, \Lambda)$ being a pair associated with a Gaussian state implies that $(A,0)$ corresponds to a pure Gaussian state.
The property $\norm{\Lambda}<1$ is explained in \cite[Remark VI.6.3]{John2021-le}.

The item \ref{item:A-Lambda-under-passive-operation} is the only statement left to prove. To this end by \eqref{eq:Gaussian-generating-function}, we have, \begin{align*}
    &\mel{e(\bar{\bu})}{\Gamma(U)\rho\Gamma(U)^{\dagger}}{e(\bv)}\\&=\mel{e(U^{\dagger}\bar{\bu})}{\rho}{e(U^{\dagger}\bv)}\\&=\mel{e(\overline{U^{T}{\bu}})}{\rho}{e(U^{\dagger}\bv)}  \\
    &=c(A,\Lambda)\exp{ (U^{T}{\bu})^T\bm{\mu}+ \bar{\bm{\mu}}^T(U^{\dagger}\bv)+(U^{T}{\bu})^TA(U^{T}{\bu})+ (U^{T}{\bu})^T\Lambda (U^{\dagger}\bv)+ (U^{\dagger}\bv)^T\bar{A}(U^{\dagger}\bv)}\\
    &=c(A,\Lambda)\exp{{\bu}^TU\bm{\mu}+ (\overline{U\bm{\mu}})^T\bv+{\bu}^TUAU^{T}{\bu}+ {\bu}^TU\Lambda U^{\dagger}\bv+ \bv^T\overline{UAU^T}\bv},
\end{align*}
proving the statement.
\end{proof}
\Frameremark{
\begin{rmks}
    \begin{enumerate}
        \item Equation~\eqref{eq:Gaussian-generating-function} written in terms of coherent states $\ket{\bv}:=\exp{-\frac{1}{2}\norm{\bv}^2}e(\bv)$  is as follows, \[\mel{\bar{\bu}}{\rho}{\bv}=c(A,\Lambda)\exp{\bu^T\bm{\mu}+ \bar{\bm{\mu}}^T\bv-\frac{1}{2}\left(\abs{\bu}^2+\abs{\bv}^2\right) + \bu^TA\bu+ \bu^T\Lambda \bv+ \bv^T\bar{A}\bv}.\]
        \item If $A$ and $\Lambda$ are real matrices, then $M(A, \Lambda)>0$ if and only if $I-\Lambda> \pm 2A$.
    \end{enumerate}
\end{rmks}
}

\section{Integral of Gaussian states}\label{sec:integral}
\Frame{
\begin{lem}\label{lem:Gaussian-average-state}Let $\rho = \rho_{\m_{\BR}, S_{\BR}}$ be a Gaussian state on $\GCn$ (see Definition \ref{defn:Gaussian-state} and the paragraph following that).
Let $\mathcal{N}=\mathcal{N}_{\bm{\mu}_{\BR},\Sigma_{\BR}}$ be the classical multivariate normal distribution on $\BR^{2n}$ with mean vector $\bm{\mu}_{\BR} = (\bm{\mu_1},\bm{\mu_2})$, $\bm{\mu}_j\in\BR^{2n}$ and covariance matrix $\Sigma_{\BR}\in M_{2n}(\BR)$.  Then \begin{equation}
    \label{eq:mean-average-state}
    \overline{\rho}:= \int\limits_{\BR^{2n}} W(\x+i\y)\rho_{\m_{\BR},S_{\BR}} W(\x+i\y)^\dagger \mathcal{N}_{\bm{\mu}_{\BR},\Sigma_{\BR}}(\dd \x\dd\y) = \bar{\rho}_{\m_{\BR}+\bm{\mu}_{\BR}, S_{\BR}+2J^T\Sigma_{\BR} J}
\end{equation}
is a Gaussian state with mean annihilation vector $\m_{\BR}+\bm{\mu}_{\BR}\in\BR^{2n}$  and covariance matrix $S_{\BR}+2J^T\Sigma_{\BR} J\in M_{2n}(\BR)$, the integral is a weak operator integral.
\end{lem}
}
\begin{proof} For any $\phi, \psi \in \GCn$ the map $(\x,\y)\mapsto \mel{\phi}{W(\x+i\y)\rho W(\x+i\y)^\dagger}{\psi}$ is continuous and therefore measurable. Furthermore, by the Cauchy--Schwartz inequality, \[\int\limits_{\BR^{2n}}\abs{\mel{\phi}{W(\x+i\y)\rho W(\x+i\y)^\dagger}{\psi}}\mathcal{N}(\dd \x \dd\y)   \leq \norm{\phi}\norm{\psi}\norm{\rho}.\]
Hence the weak operator integral on the right side of \eqref{eq:mean-average-state} is well defined. Also, by the properties of the weak operator integral, $\overline{\rho}$ is a positive trace class operator with unit trace. Write $\bu=\x+i\y$, for $\z\in \BC^n$  the quantum characteristic function of $\overline{\rho}$ at $\z$ is given by 
\begin{align*}
    \hat{\overline{\rho}}(\z)&:= \tr \overline{\rho} W(\z) = \tr \left(\left(\int_{\BC^n} W(\bu)\rho_{\m, S} W(\bu)^\dagger \mathcal{N}(\dd \bu)\right) W(\z)\right)\\
    & = \int_{\BC^n} \tr \left(\rho_{\m+\bu, S} W(\z)\right)\mathcal{N}(\dd \bu)\\
    &=  \int_{\BC^n} \hat{\rho}_{\m+\bu, S}(\z) \mathcal{N}(\dd \bu)\\
    &= \int_{\BC^n} \exp{-2i\mathscr{I} \left(\braket{\z}{\m+\bu} \right)-\mathscr{R}\left(\mel{\z}{S}{\z}\right)} \mathcal{N}(\dd \bu)\\
    &= \exp{-2i\mathscr{I} \left(\braket{\z}{\m}\right)-\mathscr{R}\left(\mel{\z}{S}{\z}\right)}\int_{\BC^n} \exp{-2i\mathscr{I} \left(\braket{\z}{\bu} \right)} \mathcal{N}(\dd \bu)\numberthis \label{eq:300}
\end{align*}
Let $\z  = \bv+i\w$ with $\bv, \w   \in \BR^n$.
We have \[\mathscr{I} \left(\braket{\z}{\bu} \right) = \bmqty{\bv^T&\w^T} J\bmqty{ \x\\ \y}. \]
Now recall that the characteristic function of the $2n$-dimensional classical normal distribution ${\mathcal{N}}_{\bm{\mu}_{\BR},\Sigma_{\BR}}$ is given by \[\hat{\mathcal{N}}_{\bm{\mu}_{\BR},\Sigma_{\BR}}(\bv,\w) = \exp{i\bmqty{\bv^T&\w^T}\bmqty{\bm{\mu}_1\\\bm{\mu}_2}-\frac{1}{2}\bmqty{\bv^T&\w^T}\Sigma_{\BR} \bmqty{\bv\\\w}},\quad \bv,\w \in \BR^n,\]
where $\bm{\mu}_j\in \BR^n,$ $\Sigma_{\BR}\in M_{2n}(\BR)$. Therefore, \begin{align*}
  \int\limits_{\BC^n} \exp{-2i\mathscr{I} \left(\braket{\z}{\bu} \right)} \mathcal{N}(\dd \bu) & = \int\limits_{\BR^{2n}} \exp{-2i\bmqty{\bv^T&\w^T} J\bmqty{ \x\\ \y}} \mathcal{N}(\dd \x \dd \y) \\
  &= \hat{\mathcal{N}}\left(-2J^T\bmqty{\bv\\\w}\right)\\
  &= \exp{-2i\bmqty{\bv^T&\w^T}J\bmqty{\bm{\mu}_1\\\bm{\mu}_2}-2\bmqty{\bv^T&\w^T}J\Sigma_{\BR} J^T\bmqty{\bv\\\w}}\\
  & = \exp{-2i\mathscr{I}\left(\braket{\z}{\bm{\mu}}\right)-2\mathscr{R}\left( \mel{\z}{J\Sigma_{\BR} J^T}{\z}\right)}.
\end{align*}
Now, continuing from \eqref{eq:300} we have, \begin{align*}
    \hat{\overline{\rho}}(\z) = \exp{-2i\mathscr{I}\left( \braket{\z}{\m + \bm{\mu}}\right)-\mathscr{R} \left(\mel{\z}{S_{\BR}+2J\Sigma_{\BR} J^T}{\z} \right)}.
\end{align*} Since $J\Sigma_{\BR} J^T=J^T\Sigma_{\BR} J$, $\overline{\rho}$ is a Gaussian state with mean vector $\m+\bm{\mu}$ and covariance matrix $S_{\BR}+2J^T\Sigma_{\BR} J$.
\end{proof}
\Frame{
\begin{thm}\label{thm:CGS}
The quantum covariance matrix $S_{\mathds{R}}$ of a classical Gaussian state $\rho = \int \ketbra{\bm{\alpha}}\mathcal{N}_{\bm{\mu}_{\BR},\Sigma_{\BR}}(\dd \bm{\alpha})$ is of the form 
\begin{align}\label{eq:cov-matrix-classical-gaussian-state}
    S_{\mathds{R}}= \frac{1}{2}I_{2n}+2J^T\Sigma_{\BR} J.
\end{align}
Consequently, an $n$-mode Gaussian state $\rho_{\m,S}$ is a classical Gaussian state  if and only if its covariance matrix $S_{\mathds{R}}$ satisfies the property 
\begin{align}
    S_{\mathds{R}}\geq \frac{1}{2}I_{2n}.
\end{align}
\end{thm}
}
\begin{proof}
Let us assume that the quantum Gaussian state $\rho= \rho_{\m,S}$ is a classical state, that is, \[\rho_{\m,S}= \int_{\BC^n}\ketbra{\bm{\alpha}}\mathcal{N}_{\bm{\mu}_{\BR},\Sigma_{\BR}}\left(\dd \bm{\alpha}\right),\]  for some $\bm{\mu}_{\BR}\in \BR^{2n}$ and $\Sigma_{\BR}
\in M_{2n}(\BR)$ with $\Sigma\geq0$. Since $\ketbra{\bm{\alpha}}=W(\bm{\alpha})\ketbra{0}W(\bm{\alpha})^{\dagger}$, and $\ketbra{0} =\rho_{0, \frac{1}{2}I_{2n}}$, we see from  Lemma~\ref{lem:Gaussian-average-state}  that  $S_{\mathds{R}}= \frac{1}{2}I_{2n}+ 2J^T\Sigma J$.
Since $\Sigma \geq 0$, this implies that $S_{\mathds{R}}\geq \frac{1}{2}I_{2n}$. 
On the other hand, if $\rho_{\m,S}$ is an $n$-mode Gaussian state with $S_{\mathds{R}}\geq\frac{1}{2}I_{2n}$, then we can choose
\begin{align}
    \Sigma_{\BR}:=\frac{1}{2}J\left(S_{\mathds{R}}-\frac{1}{2}I_{2n}\right)J^T \geq 0.
\end{align} 
Then by Lemma~\ref{lem:Gaussian-average-state}, we have
\begin{align}
    \rho_{\m,S}=\int_{\BC^n}\ketbra{\bm{\alpha}}\mathcal{N}_{\bm{\mu_{\BR}},\Sigma_{\BR}}\left(\dd \bm{\alpha}\right),
\end{align}
 where $\mathcal{N}_{\bm{\mu},\Sigma}$ is a possibly degenerate normal distribution.
\end{proof}
\end{appendices}
  
\section*{Acknowledgment}
    HKM acknowledges supports from the NSF under grant no.~2304816 and AFRL
    under agreement no.~FA8750-23-2-0031. 
    HKM thanks Vishal Singh for insightful discussions.

      \bibliographystyle{IEEEtran}
\bibliography{paperpile,sp_ref}
\end{document}